\documentclass[amssymb,aps,floatfix,floats,preprintnumbers,prl,twocolumn,superscriptaddress,groupedaddress]{revtex4}
\usepackage{rcs}
\usepackage{color,graphicx}
\usepackage{dcolumn}
\usepackage{bm}
\usepackage[normalem]{ulem}
\usepackage{chngcntr}
\usepackage[thinspace,thinqspace]{SIunits}
\usepackage{natbib}
\usepackage{amsmath}

\newcommand{\muB}{$\mu_{\textrm{B}}$}

\newcommand{\CRA}{CeRh$_{2}$As$_{2}$}
\newcommand{\CCS}{CeCu$_{2}$Si$_{2}$}
\newcommand{\Tc}{$T_{\textrm{c}}$}
\newcommand{\To}{$T_{\textrm{0}}$}

\newcommand{\TK}{$T_{\textrm{K}}$}

\newcommand{\Tm}{$T_{\textrm{m}}$}

\begin{document}
\preprint{APS/123-QED}
\title{Possible quadrupole density wave in the superconducting Kondo lattice \CRA}

\author{D. Hafner}
\email[Corresponding author:~]{daniel.hafner@cpfs.mpg.de}
\author{P. Khanenko}
\affiliation{Max Planck Institute for Chemical Physics of Solids, D-01187 Dresden, Germany}
\author{E.-O. Eljaouhari}
\affiliation{Institute for Mathematical Physics, Technische Universit\"at Braunschweig, D-38106 Braunschweig, Germany}
\affiliation{Université de Bordeaux, CNRS, LOMA, UMR 5798, 33400 Talence, France}
\author{R. K\"uchler}
\author{J. Banda}
\author{N. Bannor}
\author{T. L\"uhmann}
\author{J. F. Landaeta}
\affiliation{Max Planck Institute for Chemical Physics of Solids, D-01187 Dresden, Germany}
\author{S. Mishra}
\author{I. Sheikin}
\affiliation{Laboratoire National des Champs Magnetiques Intenses (LNCMI-EMFL), CNRS, Univ. Grenoble Alpes, 38042 Grenoble, France}
\author{E. Hassinger}
\affiliation{Max Planck Institute for Chemical Physics of Solids, D-01187 Dresden, Germany}
\affiliation{Technical University Munich, Physics department, 85748 Garching, Germany}
\author{S. Khim}
\author{C. Geibel}
\affiliation{Max Planck Institute for Chemical Physics of Solids, D-01187 Dresden, Germany}
\author{G. Zwicknagl}
\affiliation{Institute for Mathematical Physics, Technische Universit\"at Braunschweig, D-38106 Braunschweig, Germany}
\author{M. Brando}
\email[Corresponding author:~]{manuel.brando@cpfs.mpg.de}
\affiliation{Max Planck Institute for Chemical Physics of Solids, D-01187 Dresden, Germany}
\date{\today}
\begin{abstract}

\CRA\ has recently been reported to be a rare case of multi-phase unconventional superconductor~\cite{khim2021a} close to a quantum critical point (QCP). Here, we present a comprehensive study of its normal state properties and of the phase (I) below \To\ $\approx 0.4$\,K which preempts  superconductivity at \Tc\ = 0.26\,K. The $2^{nd}$-order phase transition at \To\ presents signatures in specific heat and thermal expansion, but none in magnetization and ac-susceptibility, indicating a non-magnetic origin of phase I. In addition, an upturn of the in-plane resistivity at \To\ points to a gap opening at the Fermi level in the basal plane. Thermal expansion indicates a strong positive pressure dependence of \To, $\mathrm{d}T_{\textrm{0}}/\mathrm{d}p = 1.5$\,K/GPa, in contrast to the strong negative pressure coefficient observed for magnetic order in Ce-based Kondo lattices close to a QCP. Similarly, an in-plane magnetic field shifts \To\ to higher temperatures and transforms phase I into another non-magnetic phase (II) through a $1^{st}$-order phase transition at about 9\,T. Using renormalized band structure calculations, we found that the Kondo effect (\TK\ $\approx 30$\,K) leads to substantial mixing of the excited crystalline-electric-field (CEF) states into the ground state.  This allows quadrupolar degrees of freedom in the resulting heavy bands at the Fermi level which are prone to nesting. The huge sensitivity of the quadrupole moment on hybridization together with nesting would cause an unprecedented case of phase transition into a quadrupole-density-wave (QDW) state at a temperature \To\ $\ll$ \TK, which would explain the nature of phase I and II.
\end{abstract}
\maketitle
\section{Introduction}
Unconventional superconductivity (SC) was first discovered in a heavy-fermion (HF) compound with tetragonal centrosymmetric structure, \CCS~\cite{steglich1979}. This system, as all other unconventional superconductors, shows single-phase superconductivity at ambient pressure. The only well established exception to this is UPt$_{3}$~\cite{fisher1989,bruls1990,adenwalla1990}. In this material, SC develops within a weak antiferromagnetic (AFM) order which could be responsible for phase multiplicity~\cite{sauls1994,joynt2002}.

A route to realize multi-phase SC has been theoretically proposed for spin-singlet quasi-two-dimensional multilayer superconductors in which the combination of Rashba-type spin-orbit coupling and magnetic field leads to two superconducting order parameters separated by a first order phase transition in field~\cite{yoshida2012}. This scenario seems to be realized in the recently discovered - tetragonal but locally non-centrosymmetric - HF superconductor \CRA\ (\Tc\ = 0.26\,K)~\cite{khim2021a}. Superconductivity emerges from a non-Fermi liquid (NFL) state, i.e. $C/T \propto T^{-0.6}$ and $\rho(T) \propto \sqrt{T}$, suggesting proximity to a quantum critical point (QCP) as usually observed in HF systems~\cite{monthoux2007}. Recent As-NQR experiments have detected internal fields at temperatures near \Tc, pointing to an additional AFM order~\cite{ishida2021}. Moreover, a hump in specific heat at \To\ = 0.4\,K, higher than \Tc, was observed which likely signals some non-identified order~\cite{khim2021a}. It is therefore of fundamental importance to understand the normal state properties and to identify possible phases which could compete or coexist with superconductivity in \CRA.

We report here a comprehensive study of \CRA\ based on measurements of the resistivity, specific heat, thermal expansion, magnetostriction, magnetization and torque done on several single crystals. These are complemented by renormalized band structure (RBS) calculations~\cite{zwicknagl1992,zwicknagl1993,zwicknagl2011,zwicknagl2016} using the Dirac-relativistic band structure code~\cite{christensen1984}. In the past these have been proven to agree extremely well with experimental data of strongly correlated electron systems~\cite{zwicknagl1990a,stockert2004,pfau2013}. We show that superconductivity is preempted by a non-magnetic phase (I), evidenced by signatures in specific heat and thermal expansion at \To\ $\approx 0.4$\,K in zero field and by an upturn in resistivity at \To\ suggesting a gap opening at the Fermi level. The analysis of thermal expansion and specific heat data using the Ehrenfest relation results in a strong positive pressure dependence of \To, $\mathrm{d}T_{\textrm{0}}/\mathrm{d}p = 1.5$\,K/GPa. This is in contradiction with the negative pressure coefficient expected for magnetic order in Ce-based Kondo lattices close to a magnetic QCP. With \To\ = 0.4\,K and $\mathrm{d}T_{\textrm{0}}/\mathrm{d}p = 1.5$\,K/GPa, \To\ would be suppressed at a negative pressure of only -0.27\,GPa. This result indicates that \CRA\ is close to a quantum critical point, but a very unusual one for a Ce-based Kondo lattice since the ordered state occurs on the high pressure side of the QCP. Similarly, an in-plane magnetic field $H_{\perp}$ shifts \To\ to higher $T$, in contrast to the negative field dependence expected for an antiferromagnetic order. At $\mu_{0}H_{\perp} = 9$\,T we observe a first order phase transition to a second, non-magnetic ordered state (phase II) with an even larger positive $\mathrm{d}T_{\textrm{0}}/\mathrm{d}H$.

Although our experimental techniques can not directly reveal the order parameter of both phases, our experimental and theoretical results suggest that they are multiorbital in nature despite the tetragonal crystalline electric field (CEF) having a Kramers doublet as the ground state. This is surprisingly possible here because the Kondo temperature is very close to the energy of the CEF 1$^{st}$-excited state which hybridizes strongly with the ground state allowing quadrupolar degrees of freedom in several heavy bands. Because these bands are also prone to nesting, this allows an unprecedented case of 'quadrupole density wave' (QDW) order, in analogy with the recently proposed 'metaorbital' transition~\cite{hattori2010}.

We present first all experimental results and then discuss the possible nature of the new phases I and II. Several normal state properties of \CRA\ have already been presented and discussed in Ref.~\cite{khim2021a}. Our new results are consistent with those and the same features have been observed in all samples studied here. Information about all samples and batch numbers B1, B2 and B3 can be found in the Supplemental Material (SM).

\begin{figure}[t]
	\begin{center}
		\includegraphics[width=\columnwidth]{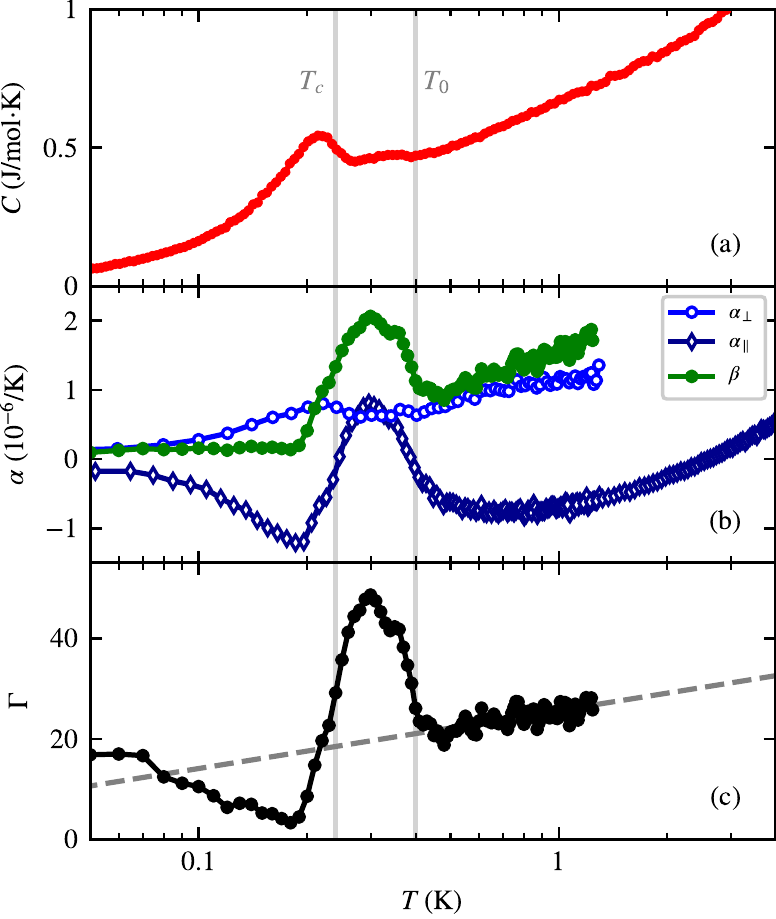}
		\caption{$T$-dependence of the (a) specific heat $C(T)$, (b) linear and volume thermal expansion coefficients, $\alpha_{\perp}(T)$, $\alpha_{\parallel}(T)$, $\beta(T)$, and the (c) Gr\"uneisen parameter $\Gamma(T)$. The vertical grey lines indicate the onset of the transition at \To\ and \Tc. Measurements done on samples from B3 ($l_{\perp} = 2.24$\,mm and $l_{\parallel} = 0.73$\,mm).}
		\label{fig1}
	\end{center}
\end{figure}

A summary of the zero-field data on a sample from B3 is presented in Fig.~\ref{fig1}. The specific heat $C(T)$ versus $T$ shows a weak but clear hump starting at \To\ = 0.4\,K and a second order superconducting phase transition at \Tc\ = 0.24\,K similar to what was observed in Ref.~\cite{khim2021a}. 
Because of the weak signature at \To\, we employed thermal expansion measurements as another thermodynamic probe to confirm the presence of a phase transition. We have measured the thermal expansion coefficients along the $c$-axis $\alpha_{\parallel} = 1/l_{\parallel} \cdot \partial l/ \partial T$ and in the plane $\alpha_{\perp} = 1/l_{\perp} \cdot \partial l/ \partial T$. $\alpha_{\perp}(T)$ is positive in the whole $T$-range whereas $\alpha_{\parallel}(T)$ changes sign near \To\ and \Tc. The volume thermal expansion coefficient $\beta(T) = \alpha_{\parallel} + 2\alpha_{\perp}$ is, however, overall positive as expected for a Ce-based Kondo lattice system at $T <$ \TK~\cite{devisser1990}, where \TK\ is the Kondo temperature estimated to be about 30\,K for \CRA~\cite{khim2021a}. This is because $\beta(T)=-k_{T}\left(\partial S/\partial p\right)_{T}$ ($k_{T}$ = compressibility) and in Ce-based Kondo-lattice (KL) systems at $T <$ \TK, $\partial S/\partial p$ is negative because increasing pressure enhances \TK\ which results in a decrease of $S$ at a given $T$. On the contrary, below a putative magnetic order at a temperature \Tm, $\beta(T)$ is expected to be negative since \Tm\ decreases with increasing pressure in Ce-based Kondo lattices close to a magnetic QCP~\cite{doniach1977,cornelius1995}.

Both coefficients $\alpha_{\parallel}(T)$ and $\alpha_{\perp}(T)$ show clear jumps at \Tc\ as expected for a 2$^{nd}$-order phase transition. The jump is positive in $\alpha_{\perp}(T)$ and negative in $\alpha_{\parallel}(T)$, as sometimes observed in other HF superconductors like CeIrIn$_{5}$~\cite{oeschler2003}. However, across \To, $\alpha_{\parallel}(T)$ exhibits a clear jump, but only a weak feature could be detected in $\alpha_{\perp}(T)$. The total jump in $\beta(T)$ is as large as that at \Tc\ implying that we also have a 2$^{nd}$-order phase transition at \To. The strong anisotropy in the thermal expansion coefficients is very rare and implies that \To\ is suppressed for a uniaxial stress only along the $c$-axis. Using the Ehrenfest relation $\mathrm{d}T_{\textrm{c}}/\mathrm{d}p = V_{mol} T_{\textrm{c}} \left(\Delta \beta/\Delta C\right)$ ($V_{m}$ = molar volume) we obtain $\mathrm{d}T_{\textrm{c}}/\mathrm{d}p = -0.21$\,K/GPa and $\mathrm{d}T_{\textrm{0}}/\mathrm{d}p = 1.5$\,K/GPa (see Fig.~S4 in SM). So, the pressure dependencies have opposite sign, as commonly seen in KL systems close to a QCP at which magnetic order is replaced by a superconducting state. However, in strong contrast to all known Ce-based QCP systems, the $p$-dependence of \To\ is positive, which implies that the order associated with \To\ is stabilized by pressure, not suppressed as expected and observed for AFM order. With \To\ = 0.4\,K and $\mathrm{d}T_{\textrm{0}}/\mathrm{d}p = 1.5$\,K/GPa, \To\ should completely be suppressed at a negative pressure of -0.27\,GPa, while \Tc\ should be only slightly enhanced by about 57\,mK at this negative pressure. Thus, this analysis based on the Ehrenfest relation indicates that in the \Tc$(p)$ - \To$(p)$ phase diagram of \CRA\ the $p$-dependence is inverse compared to presently known Ce-based KL superconductors. This reversed $p$-dependence is a strong indication that the order associated with \To\ can not be a standard AFM order, in agreement with the lack of signature in ac-susceptibility at \To (see Ref.~\cite{khim2021a} and Fig.~S3 of the SM).

A fundamental quantity in KL systems is the Gr\"uneisen parameter $\Gamma(T) = k_{T} \cdot V_{mol} \cdot \beta(T)/C(T)$. The ratio between thermal expansion and specific heat reflects the pressure dependence of the characteristic energy of the system~\cite{zhu2003}. Since in KL systems close to a QCP the magnetic ordering temperature as well as \TK\ are strongly pressure sensitive, their Gr\"uneisen parameter is usually very large and shows a strong $T$-dependence. This can thus be used as a powerful analyzing tool~\cite{zhu2003,kuechler2003,steppke2013,watanabe2019}. $\Gamma(T)$ of \CRA\ is shown in Fig.~\ref{fig1}c. As expected, it is quite large with a strong and complex $T$ dependence. The latter property is certainly connected with the two competing orders at \To\ and \Tc. Remarkably, $\Gamma(T)$ is positive in the whole $T$-range studied here, as in paramagnetic Ce-based KL systems. This implies that the dominant interaction has a positive $p$-dependence, suggesting that the dominant interaction that governs the $p$-dependent properties of \CRA\ might still be the Kondo exchange, i.e. the hybridization between $4f$ and conduction electrons, as in standard KL systems, but with a different nature of the ordering at \To. Thus combining the analysis of the Gr\"uneisen ratio with the analysis of the Ehrenfest relation indicates that increasing the $4f$-hybridization stabilizes the ordering at \To. A further notable result is that the $T$-dependence of $\Gamma(T)$ above \To\ does not show any evidence for the divergence expected at a QCP. Instead it seems to decrease logarithmically with decreasing $T$. Only well below \Tc\ it seems to increase with decreasing $T$ reaching values of about 20, sometimes observed in other HF superconductors~\cite{devisser1990,oeschler2003}. Since other properties as, e.g., the NFL behavior seen in the specific heat supports the closeness to a QCP, the absence of a well-defined divergence in $\Gamma(T)$ and its complex $T$-dependence might be the result of multiple competing phenomena.
\begin{figure}[t]
	\begin{center}
		\includegraphics[width=\columnwidth]{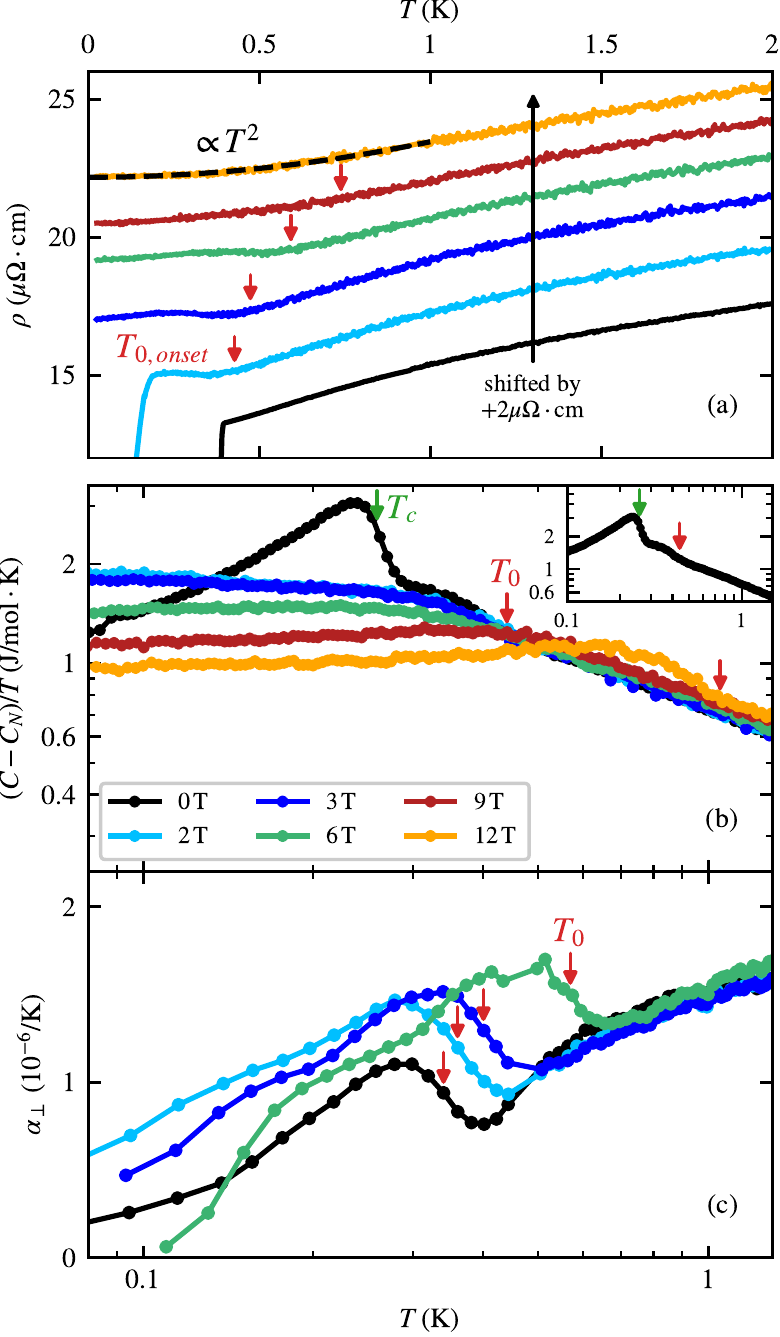}
		\caption{$T$-dependence of the (a) resistivity $\rho(T)$, (b) electronic specific heat coefficient $(C - C_{N})/T$ and (c) linear thermal expansion coefficient $\alpha_{\perp}(T)$. $C_{N}$ is the nuclear contribution to the specific heat. Red and green arrows indicate \Tc\ and \To, respectively. Measurements done on samples from batches B1 and B2 with $H_{\perp}$ perpendicular to the $c$-axis.}
		\label{fig2}
	\end{center}
\end{figure}

We have then investigated the effect of an in-plane magnetic field $H_{\perp}$ on the properties of \CRA\ in order to gain more information about the phase associated with \To. This is summarized in Fig.~\ref{fig2} in which we plotted the resistivity $\rho(T)$, the electronic specific heat coefficient $(C(T)-C_{N}(T))/T$ and the in-plane thermal expansion coefficient $\alpha_{\perp}(T)$ at different $\mu_{0}H_{\perp}$. We first analyze $\rho(T)$ measured on the same sample of Ref.~\cite{khim2021a} and shown in Fig.~\ref{fig2}a. As explained in Ref.~\cite{khim2021a}, the drop due to SC in $\rho(T)$ is found at a higher temperature (0.39\,K) than the \Tc\ detected in specific heat or susceptibility. This is ascribed to inhomogeneity in the material as known from other HF systems~\cite{hassinger2008,bachmann2019}. This effect prevents us from seeing which feature $\rho(T)$ may exhibit at \To\ = 0.4\,K. Fortunately, since \To\ increases with increasing $H_{\perp}$ while superconductivity is suppressed at $\mu_{0}H_{c2} \approx 2$\,T (cf. Fig.~\ref{fig4}), we can detect and follow the signature at \To\ for $\mu_{0}H_{\perp} \geq 2$\,T. In Fig.~\ref{fig2}a a very clear kink is seen at \To\ = 0.4\,K in $\rho(T)$ measured with $\mu_{0}H_{\perp} = 2$\,T. While $\rho(T) \propto \sqrt{T}$ above \To\ (see Fig.~S2 in SM), it shows a slight upturn below \To, also visible at higher fields of 4\,T and 6\,T. At 4\,T the value of $\rho$(0.1\,K) is increased by over 20\% compared to the value extrapolated from $\rho(T)$ above \To. This kind of signature is generally observed at spin density wave (SDW)~\cite{hamann2019} or charge density wave (CDW)~\cite{gruner2017} phase transitions and indicates the opening of a gap at the Fermi level caused by nesting. Since the current was applied in the basal plane of the tetragonal structure, this implies that the propagation vector of the order parameter below \To\ has a component within the basal plane.

\begin{figure}[t]
	\begin{center}
		\includegraphics[width=\columnwidth]{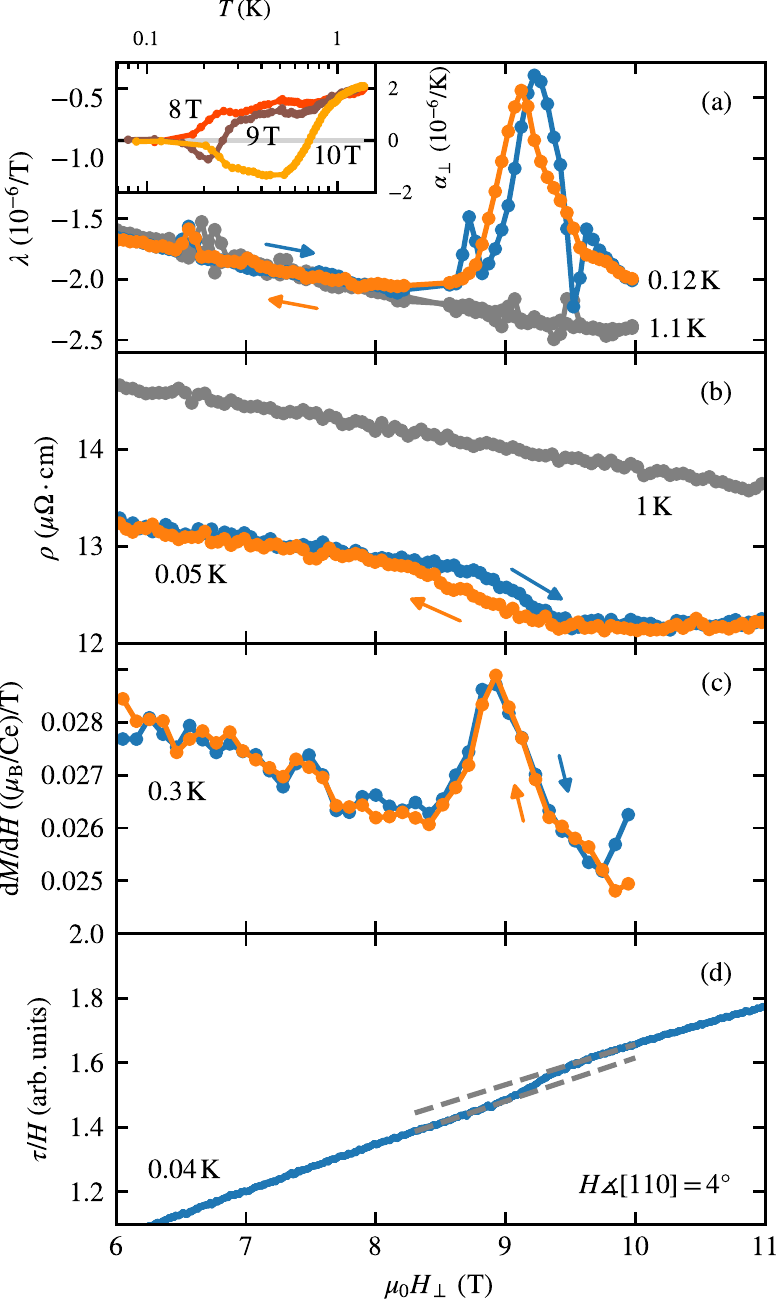}
		\caption{Field-dependence of (a) the magnetostriction coefficient $\lambda(H)$, (b) resistivity $\rho(H)$, (c) field derivative of magnetization $\mathrm{d}M(H)/\mathrm{d}H$ and (d) torque $\tau(H)/H$ with field $H_{\perp}$ applied within the basal plane. The inset of panel (a) shows the $T$-dependence of the in-plane thermal expansion coefficient $\alpha_{\perp}(T)$ for fields $\mu_{0}H_{\perp} \geq 8$\,T. Measurements done on samples from B2.}
		\label{fig3}
	\end{center}
\end{figure}

\begin{figure}[t]
	\begin{center}
		\includegraphics[width=\columnwidth]{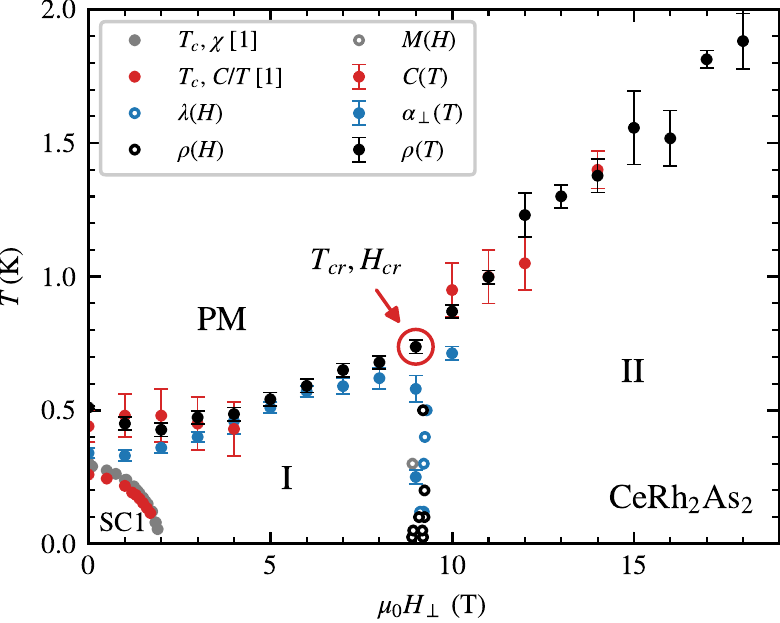}
		\caption{$T - H_{\perp}$ phase diagram of \CRA. The points on the phase diagram have been extracted from measurements performed on samples from B1 and B2 which show the same \Tc\ and \To. The point $(T_{\textrm{cr}}=0.7\,\textrm{K},H_{\textrm{cr}}=9\,\textrm{T})$ is the tricritical point at the end of the first order boundary which separates phase I from phase II. If error bars are not displayed, they are smaller than the symbols. The points for the SC $T_c$ were first published in \cite{khim2021a}.}
		\label{fig4}
	\end{center}
\end{figure}

The upturn in $\rho(T)$ disappears for $\mu_0H_{\perp} \geq 10$\,T after which we observe a broad crossover from $\rho(T) \propto \sqrt{T}$ into a Fermi-liquid-like behavior $\rho(T) \propto T^{2}$ (see fit on 12\,T curve in Fig.~\ref{fig2}a). The crossover temperatures coincide with phase transition temperatures observed in thermal expansion and specific heat (cf. Fig.~\ref{fig4}). In contrast, the evolution of the shape of the anomalies in $\rho(T)$ and $C(T)/T$ with increasing field differs remarkably. In $\rho(T)$ the kink at \To\ is most pronounced in the field range $2 \leq \mu_{0}H_{\perp} \leq 8$\,T. On the other hand, the small kink discernible in $C(T)/T$ at \To\ in zero field (Fig. \ref{fig2}b) becomes weaker with increasing field so that in the range $3 \leq \mu_{0}H_{\perp} \leq 10$\,T only a broad hump is visible. But surprisingly, at even higher fields it sharpens again and for $\mu_0H_{\perp} \geq 10$\,T becomes a well-defined maximum in $C(T)/T$. This is corroborated by the behavior of $\alpha_{\perp}(T)$ in fields $H_{\perp} \geq H_{c2}$, shown in Fig.~\ref{fig2}c. The step-like increase of $\alpha_{\perp}(T)$ at \To\ is well pronounced at these fields and increases with increasing field remaining positive for $H_{\perp} \leq 8$\,T. But for $H_{\perp} > 8$\,T it shows a strong drop to negative values (see inset of Fig.~\ref{fig3}a), as it is expected, e.g., for a second order phase transition from a paramagnetic (PM) state into a magnetically ordered phase, like in the SDW phase of \CCS~\cite{stockert2004}. This is a first indication for a magnetic field induced phase transition at a field of about 9\,T. 

For this reason we have measured the $H_{\perp}$-dependence of the magnetostriction coefficient $\lambda(H) = 1/\mu_0 l_0 \cdot \partial l/ \partial H$, resistivity $\rho(H)$, magnetization $M(H)$ and torque $\bm{\tau}(H) = \bm{M} \times \bm{H}$. The results are summarized in Fig.~\ref{fig3}. We clearly observe a first order phase transition at $\mu_{0}H_{\textrm{cr}} = 9$\,T in form of a large peak in $\lambda(H)$ ($\Delta \lambda \approx 2.0 \cdot 10^{-6}/$T), a drop in $\rho(H)$, both with clear hysteresis, and a very small upturn in $M(H)$ ($\Delta M \approx 0.003$\,\muB/Ce) which is better identified in the derivative $\mathrm{d}M(H)/\mathrm{d}H$. All features vanish at $T \geq 1$\,K (gray points). The clear sign change in $\alpha_{\perp}(T)$ at the transition temperature for $H_{\perp} \geq H_{\textrm{cr}}$ (inset of Fig.~\ref{fig3}a) and the peak in $\lambda(H)$ at $H_{\textrm{cr}}$ definitively rule out a Lifshitz-like transition~\cite{pfau2017} or a kind of metamagnetic transition~\cite{weickert2010,deppe2012} and could indicate a transition into a magnetically (purely dipolar) ordered state. However, the very weak features in $M(H)$, $\tau(H)$ and $\rho(H)$ point to a phase transition into another multipolar phase possibly with induced weak dipolar moment. This is corroborated by the lack of features in $\chi(T)$ for $H_{\perp} \geq H_{\textrm{cr}}$ (see Fig.~S3 of the SM).  The field-induced transition is first order as expected when there is a change from one order parameter to another. 

We present now the $T - H_{\perp}$ phase diagram in Fig.~\ref{fig4}. It summarizes our experimental results on samples from batches B1 and B2, together with those (superconducting phase SC1) already presented in Ref.~\cite{khim2021a}. We omitted some \To-points from $C(T)$ between 5 and 9\,T since the transition is too broad to allow us to identify a well defined transition temperature. In zero field \CRA\ undergoes a phase transition at \To\ $\simeq 0.4$\,K into a non-magnetic phase I and then into a superconducting phase SC1 at \Tc\ = 0.26\,K. We could not find signatures of any other transition. \To\ increases with an in-plane field $H_{\perp}$ at a rate of about 0.04\,K/T up to the critical field $\mu_{0}H_{\textrm{cr}} = 9$\,T after which it increases much more steeply at about 0.14\,K/T. At this critical field a vertical, first order phase boundary separates phase I from another phase II whose primary order parameter is also non magnetic. Thus this line corresponds to a sort of 'metamultipolar' transition. The point $(T_{\textrm{cr}}=0.7\,\textrm{K},H_{\textrm{cr}}=9\,\textrm{T})$ at which this vertical phase boundary meets the \To$(H)$ boundary corresponds to a tricritical point.

We discuss now the possible nature of the two phases I and II, and their possible relation to SC. Specific heat and especially thermal expansion data provide clear evidence for the presence of a bulk thermodynamic second order phase transition at \To. In such a Kondo lattice one would usually assume \To\ to indicate AFM order. However, this is in disagreement with several experimental results: no related features in the susceptibility and magnetization (cf. Ref.~\cite{khim2021a} and Fig.~S3 of the SM), and the positive pressure and magnetic field dependence of \To. The latter property immediately suggests a quadrupolar or a more complex multipolar order. On the other hand, this positive field dependence of \To\ contradicts a valence transition as well as a structural CDW transition, which would anyway be very unlikely at such a low \To.

Multipolar ordered states are observed in Ce-based systems with cubic CEF symmetry, arising from the quadrupolar degrees of freedom of a quartet ground state as in CeB$_{6}$~\cite{effantin1985,jang2017} or Ce$_{3}$Pd$_{20}$Ge$_{6}$~\cite{kitagawa1996}.  But in \CRA\ the tetragonal CEF splits the $J = 5/2$ Hund's rule ground state of Ce$^{3+}$ into three Kramers doublets $\left|\Gamma\left(i\right),\tau\right\rangle $, $i=0,1,2$, that have only a spin degree of freedom $\tau=\pm$.  A determination of the CEF scheme (see SM) puts the first and the second excited CEF doublets at $\Delta_{1}\simeq30$\,K and $\Delta_{2}\simeq180$\,K above the CEF ground state doublet, respectively. Thus at \To\ $\simeq 0.4$\,K only the CEF ground state is populated, which does not bear a quadrupolar degree of freedom, and therefore quadrupolar or multipolar order should not be possible.

A quadrupolar order arising unexpectedly from a similar tetragonal CEF was found in YbRu$_{2}$Ge$_{2}$~\cite{jeevan2006} at \To\ = 10.2\,K. In this compound, the small level splitting $\Delta_{1} \approx 12$\,K $\approx$ \To\ allows effective intersite interactions between the multipole moments of the CEF levels to induce a quadrupolar order parameter~\cite{takimoto2008}. In such a case, the ordering temperature can hardly be below  $\Delta_{1}/2$. While \CRA\ has a similar $\Delta_{1}\simeq30$\,K, $T_0\simeq0.4$\,K is almost two orders of magnitude lower than that, making this scenario very unlikely for our case.

\begin{figure}[ht!]
\begin{center}
\includegraphics[width=\columnwidth]{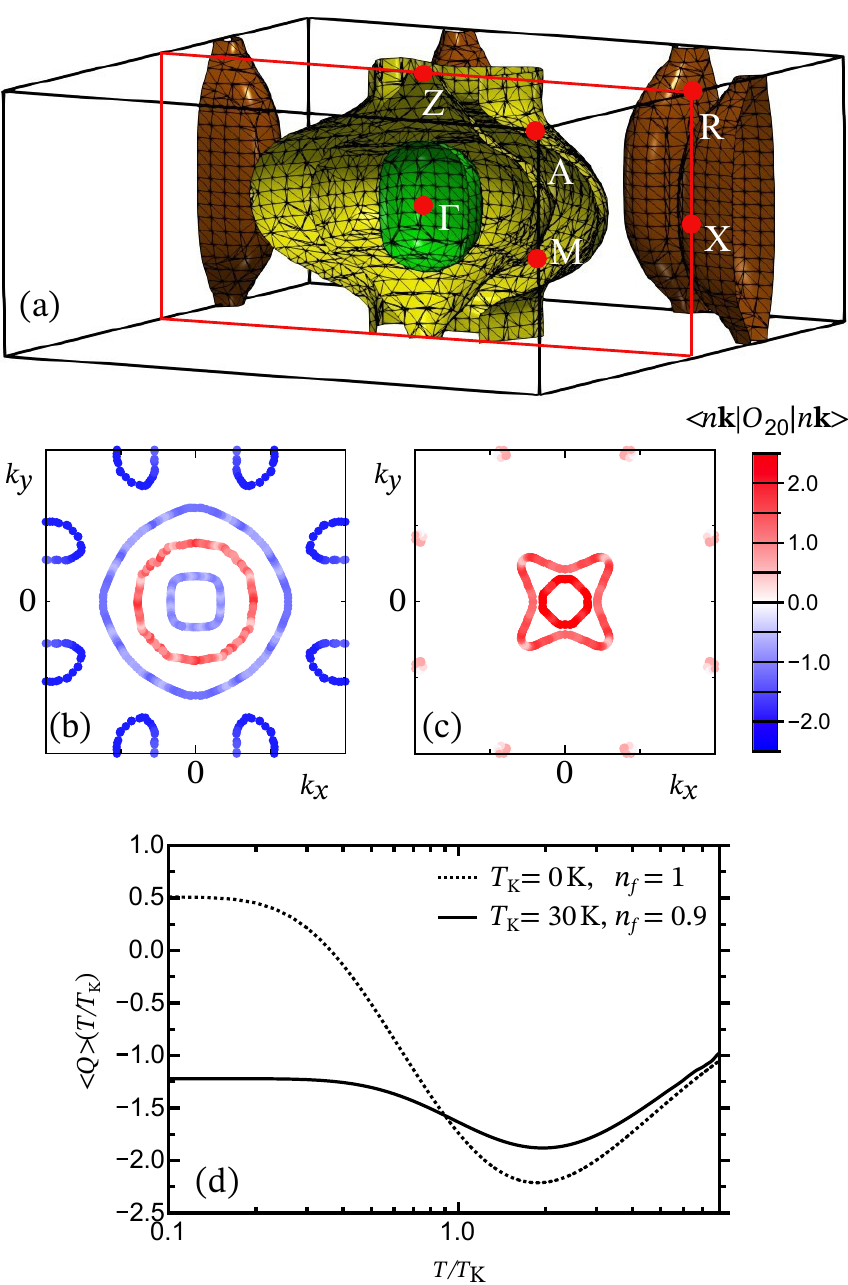}
\caption{Interplay of Kondo effect and quadrupolar degrees of freedom for the low-energy excitations in \CRA. In the figures, $B_{44} > 0$ has been chosen for CEF scheme (see SM). Assuming $B_{44} < 0$ leads to qualitatively similar results (see SM). (a) Quasiparticle FS as given by the RBS method for $B_{44} > 0$. Only half of the Brillouin zone is shown to reveal the FS structure. The characteristic features are two strongly corrugated cylinders (yellow) along $\Gamma-Z$, a $\Gamma$-centered closed surface (green) as well as corrugated cylinders (brown) along axes parallel to $A-M$. The yellow FS-sheets belong to one band and the green and brown FS-sheets belong to another band. Several critical points in the vicinity of the Fermi energy $E_{F}$ indicate the possibility of Lifshitz transitions. The sign of $B_{44}$ affects the topology only weakly. (b),(c) Pronounced nesting features are seen from the intersections of the FS with the basal plane $k_{z}=0$ and the top plane $k_{z}=\pi/c$. The ${\bf k}$-dependent expectation values of $\left\langle n{\bf k}\left|3J_{z}^{2}-{\bf J}^{2}\right|n{\bf k}\right\rangle =\left\langle n{\bf k}\left|O_{20}\right|n{\bf k}\right\rangle$ highlight that quadrupolar degrees of freedom should play an important role in this material. (d) Variation with temperature of the quadrupole moment of the single-impurity Ce $4f$ shell in the Kondo regime (\TK\ = 30\,K, $n_{f} = 0.9$, full line) differs both qualitatively and quantitatively from its counterpart in the local moment regime (\TK\ = 0, $n_{f} = 1$, dotted line).}
\label{fig5}
\end{center}
\end{figure}

However, there is another energy scale which plays a fundamental role in \CRA: The Kondo temperature \TK\ $\approx 30$\,K~\cite{khim2021a}, which has been shown to affect the quadrupole moment of CEF-split 4$f$-states~\cite{zwicknagl1988,zwicknagl1990,yamada2019,tazai2019}. As $\Delta_{1}\simeq T_{K}$, the formation of the heavy Fermi liquid state leads to a significant admixture of excited CEF states to the low-energy states. This is revealed by our RBS calculations whose main results are shown in Fig.~\ref{fig5}. Of particular interest is the expectation value of the quadrupole moment $Q$ which depends on the differences in the population of the Kramers doublets. The influence of the Kondo effect can be seen in Fig.~\ref{fig5}d in which we compare the variation with temperature $T$ of $Q\left(T\right)=\sum_{\Gamma\left(i\right),\tau}\left\langle \Gamma\left(i\right),\tau\left|3J_{z}^{2}-{\bf J}^{2}\right|\Gamma\left(i\right),\tau\right\rangle n_{f\Gamma\left(i\right)\tau}\left(T\right)$ for a Ce impurity in a dilute magnetic alloy in the presence of the Kondo effect (\TK\ = 30\,K, $n_{f} = 0.9$, full line) and in the local moment regime (\TK\ = 0, $n_{f} = 1$, dotted line). The calculations have employed a simple approximation to the Non-Crossing Approximation~\cite{zwicknagl1988,zwicknagl1990} for the Anderson model impurity Hamiltonian in the limit of large Coulomb repulsion $U$. The choice of the low-temperature $4f$ valence in the Kondo regime, $n_{f}\left(T\to0\right)$, enters mainly as an overall scale. In the low-temperature limit $T \rightarrow 0$, the Kondo effect strongly affects both the sign and the magnitude of $Q\left(T\to0\right)$. Thus, due to the strong mixing of the excited CEF doublets into the ground state, the quadrupole moment becomes a property which even at $T = 0$ is very sensitive to change in the hybridization. That opens a door for quadrupolar or multipolar ordering. 

The characteristic admixture of excited CEF states is also evident in the coherent heavy quasiparticles $\left|n{\bf k}\right\rangle $ of band index $n$ and wave number ${\bf k}$ which form in a periodic Kondo lattice. The expectation values $\left\langle n{\bf k}\left|3J_{z}^{2}-{\bf J}^{2}\right|n{\bf k}\right\rangle$ in the states at the Fermi surface (FS) are strongly ${\bf k}$-dependent as shown in Fig.~\ref{fig5}b and c. This demonstrates that quadrupolar degrees of freedom play an important role for the ground state properties of \CRA. Moreover, pronounced nesting features are seen from the intersections of the FS with the basal plane $k_{z}=0$ (Fig.~\ref{fig5}b) and the top plane $k_{z}=\pi/c$ (Fig.~\ref{fig5}c). So, if nesting occurs then the associated quadrupolar degrees of freedom will be affected. This property could link the observed opening of a gap in the plane at \To\ with the transition observed in thermodynamic probes, as discussed above. 

The results of these theoretical calculations, i.e. the strong change in the quadrupole moment induced by the Kondo exchange, the large variation of the expectation value for the quadrupole moment across the Fermi surface, and the strong nesting properties of the Fermi surfaces suggest a possible scenario which reconciles all observed properties: a quadrupolar or multipolar density wave (QDW) based on itinerant heavy $4f$ electrons. This is a very unusual scenario. To the best of our knowledge it has never been proposed before. It requires very specific properties of the $4f$ electrons, which however are all fulfilled in the case of \CRA: a significant strength of the Kondo interaction, an energy splitting to the first excited CEF doublet of the same size of the Kondo scale, a strong variability of the expectation of the quadrupole moment across the Fermi surface, and nesting features of the Fermi surface. Within this scenario the unexpected and very unusual positive pressure dependence of \To\ can easily be explained. Without doubt, pressure shall enhance the hybridization between $4f$ and conduction electrons in \CRA\ as it does in all known Ce-based HF systems, and is suggested by the positive Gr\"uneisen parameter. That should increase the change in the quadrupole moment at $T = 0$ as well as its sensitivity to hybridization, thus promoting an itinerant QDW. On the other hand, the positive field dependence of \To\ is likely due to the same mechanism as in standard, local quadrupolar ordered systems: There, the quadrupolar order is stabilized by a field-induced magnetic moment. 

So we might have in \CRA\ an uncommon quadrupolar order of itinerant $4f$ states induced by a strong $4f$-conduction electron hybridization which mixes excited CEF states into the ground state. In standard Ce- or Yb-based Kondo lattices usually only the CEF ground state is relevant and taken into account for the properties at low $T$. In contrast, in this new itinerant QDW state the excited CEF levels are crucial. Thus, it presents some analogies to the recently proposed 'metaorbital' scenario~\cite{hattori2010}, in which at higher pressures a strongly correlated $4f$ state is formed not out of the CEF ground state, but with the excited CEF state when the hybridization of this excited state is much stronger than that of the CEF ground state. We note that an unusual ordered state, with a very unusual phase diagram presenting some analogy to that of \CRA, and where very likely excited CEF states are involved, has recently been observed in the tetragonal Kondo lattice CeCoSi~\cite{lengyel2013}. As for \CRA, the Ce-site does not present inversion symmetry, while the overall structure is centrosymmetric. At zero pressure CeCoSi exhibits AFM order at 9\,K, but under pressure a new ordered state appears with an ordering temperature $T^{*}$ shooting up to 38\,K at a pressure of merely 1.2\,GPa. Increasing pressure further to 2.2\,GPa suppresses completely all ordered states, resulting in a strongly hybridized paramagnetic Kondo lattice. Recent NQR and NMR results proved that the state forming at $T^{*}$ cannot be a dipolar order, instead a yet unidentified unusual kind of multipolar order was proposed~\cite{manago2021}. However, inelastic neutron results and specific heat data put the first excited CEF doublet at 125\,K above the ground state doublet, making a standard quadrupolar order at 38\,K impossible~\cite{nikitin2020}. Therefore, a further interaction is required to stabilize multipolar order under such conditions. Since this unusual phase appears under pressure only slightly before the onset of the highly hybridized paramagnetic state, hybridization is expected to play an important role.

Concerning the nature of the high field phase II in \CRA, some insights can be gained from experimental observations: The large positive slope $\mathrm{d}T_0/\mathrm{d}H$ precludes the main order parameter to be a dipolar AFM one. The absence of any anomaly in the ac-susceptibility (Fig. S3 in SM), and the observation that the signature of the transition becomes sharper with increasing field exclude a dipolar FM order parameter. On the other hand, the large positive slope $\mathrm{d}T_0/\mathrm{d}H$ and the vertical first order phase boundary at $\mu_0H_{cr}=9\,\mathrm{T}$ with a large anomaly in the magnetostriction are reminiscent of observation in quadrupolar ordered systems. In these systems the field-induced magnetic moment couples to multipolar degrees of freedom, inducing new multipolar states which are stabilized by an increasing magnetic field. Similar experimental features have been observed, e.g., in Ce$_{0.5}$La$_{0.5}$B$_{6}$: In this compound, the crossover to phase IV is characterized by a large maximum in $C/T$ in zero field (similar to what we see at \To\ in \CRA), and a magnetic field $H \parallel [110]$ induces a very sharp phase transition from the AF octupolar (phase IV) into the AF quadrupolar phase (phase II)~\cite{jang2017}. As in \CRA,  the transition temperature in the high-field phase increases steeply with increasing field and all magnetic features observed at the critical field are weak. This strong similarity obviously suggests that the transition at $\mu_0H_{cr}=9\,\mathrm{T}$ in \CRA\ marks the change to a different multipolar order parameter stabilized by the field induced magnetic moment. In contrast to the situation in CeB$_6$, in \CRA\ the ordering multipolar degree of freedom must be a highly itinerant $4f$ one.

To conclude, the present study confirms that the superconducting state in the locally non-centrosymmetric Kondo lattice \CRA\ is preceded by an ordered state, phase I, with a transition temperature $T_0=0.4\,\mathrm{K}$ above $T_c = 0.26\,\mathrm{K}$. Phase I changes into a second phase II through a first order phase transition at an in-plane magnetic field of 9\,T. Both phases present unusual features which are incompatible with the usually expected dipolar magnetic order. Instead, all experimental results, as well as RBS calculations point to a unique case of 'quadrupole density wave' order in phase I and II. This is made possible in \CRA\ by a comparatively large \TK\ and a small crystal field splitting, which results in a very strong mixing of excited CEF doublets into the ground state doublet. As a result the heavy bands acquire quadrupolar degrees of freedom, which together with strong nesting properties open the way for quadrupolar density waves. The Pauli limited superconducting phase SC1 is entirely included within phase I (see Fig.~\ref{fig4}). If the order parameter of phase I was originating from completely localized $4f$ states, then one would expect coexistence of SC and phase I with only limited interference, as e.g. in the rare earth borocarbides~\cite{andersen2006}. But there is no doubt that the 4$f$ electrons are strongly delocalized in phase I and II of \CRA. Then one would expect a strong interference between SC and phase I if, e.g., the SC gap and the gap of phase I opens on the same sheet of the Fermi surface. However, right now there is no clear experimental evidence how the SC state and phase I are interacting.
\begin{acknowledgments}
We are indebted to D. Agterberg, H.-U. Desgranges, K. Ishida, A. P. Mackenzie, H. Rosner, O. Stockert and P. Thalmeier for useful discussions. We would like to thank the Deutsche Forschungsgemeinschaft (DFG) from Project Nos. BR 4110/1-1 and KU 3287/1-1 for financial support. This work has also been supported by the joint Agence National de la Recherche (ANR) and DFG program Fermi-NESt through grants GE602/4-1 (CG) and ZW77/5-1 (GZ).
\end{acknowledgments}
\newpage
\noindent
\textbf{\Large Supplemental Material}
\renewcommand{\thefigure}{S\arabic{figure}}
\setcounter{figure}{0} 
\subsection{Samples}
\vspace{-0.2cm}
We have measured samples from three different batches: B1 (batch 87334, \Tc\ = 0.26\,K), B2 (batch 87444, \Tc\ = 0.26\,K) and B3 (batch 87479, \Tc\ = 0.24\,K). Samples from the same batch show identical properties. Samples from B1 and B2 have also been used in Ref.~\cite{khim2021a}. Samples from a new batch B3 have been considered here, although they show a slightly lower \Tc\ = 0.24\,K, because of their larger size along both the $a$-axis ($l_{\perp}$) and $c$-axis ($l_{\parallel}$), which guarantees a good resolution in thermal expansion. A comparison of the specific heat of samples from all batches is presented in Fig.~\ref{figS1}.
\subsection{Specific heat}
\begin{figure}[ht!]
	\begin{center}
		\includegraphics[width=\columnwidth]{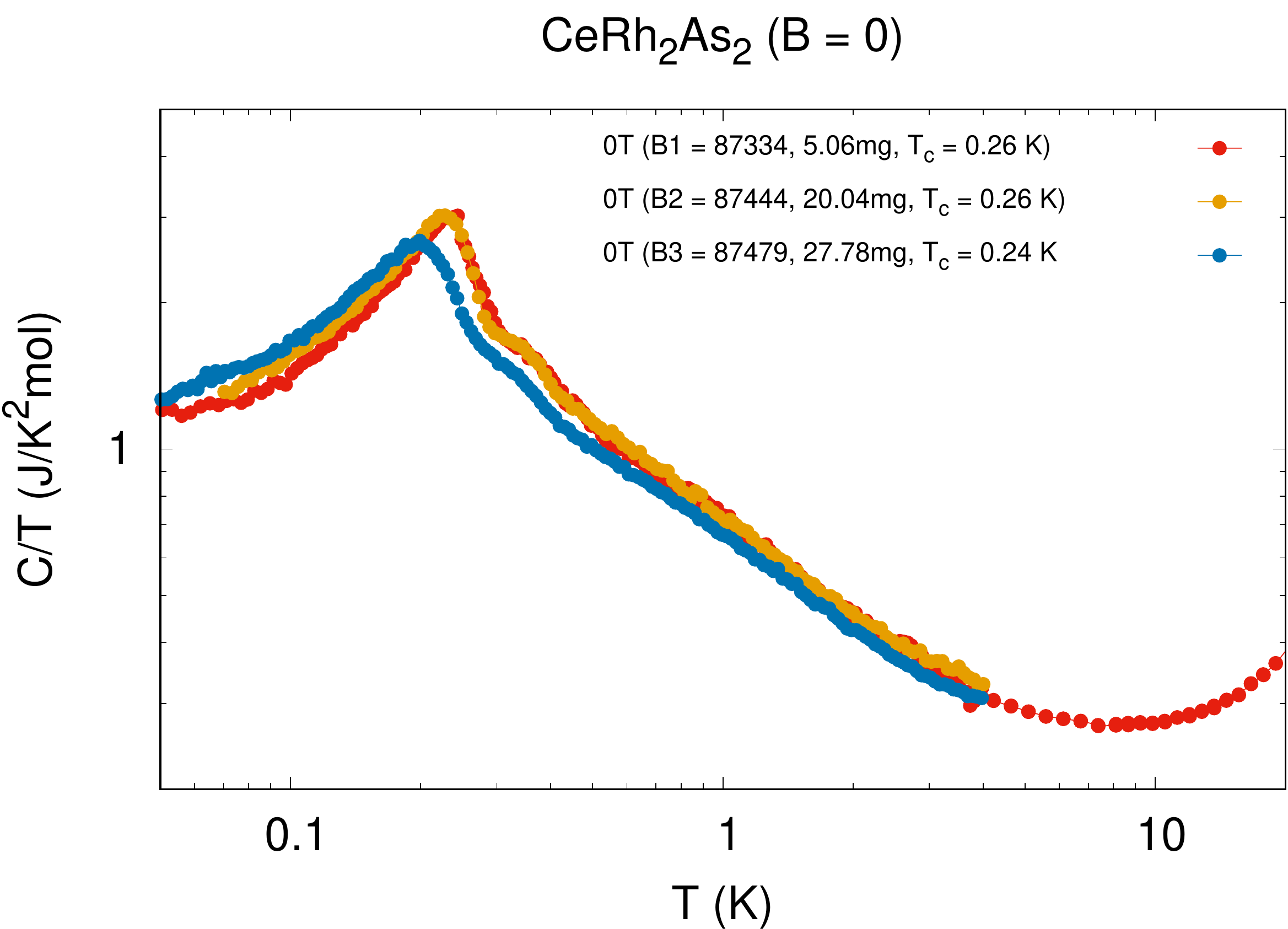}
		\caption{Temperature dependence of the specific heat for different batches of \CRA}
		\label{figS1}
	\end{center}
\end{figure}

\vspace{-0.2cm}
The onset of the phases I and II can be identified from the hump in the temperature dependence of the specific heat, as explained in text and Fig.~1 and 2 of the main article. At field strengths between 5 and 9\,T, however, this broadens so that the position of the maximum could not be significantly resolved. Therefore, we did not include points for these fields in the phase diagram of Fig.~4. Recent specific heat measurements on a much larger sample from batch B3 with a slightly lower $T_c = 0.24$\,K (see Fig.~\ref{figS1}) were able to confirm the presence of this feature in this field region. These points were however not included in the phase diagram in Fig.~4 of the main article, since this batch deviates from the other batches with a slightly lower $T_c$ and $T_0$, as it can be seen in Fig.~\ref{figS1}. The specific heat was measured with a semiadiabatic heat pulse technique~\cite{wilhelm2004}.
\subsection{Thermal expansion}
\vspace{-0.2cm}
In thermal expansion, the transition into phases I is a clear step as shown in Fig.~2 of the main article. We took the midpoint of the step as the transition temperature.
The thermal expansion was measured with a compact and miniaturized high resolution capacitance dilatometer~\cite{kuechler2012}.
\subsection{Resistivity}
\begin{figure}[ht!]
	\begin{center}
		\includegraphics[width=\columnwidth]{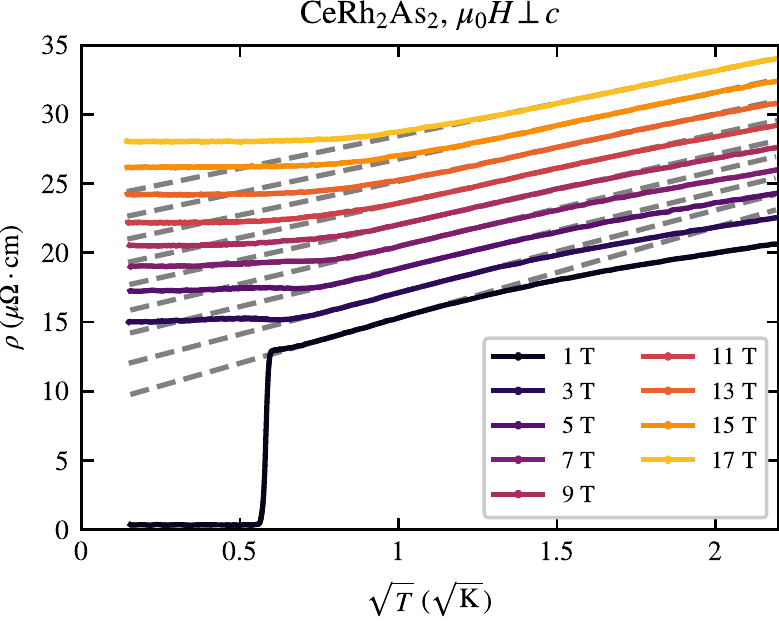}
		\caption{Temperature dependence of the resistivity $\rho(T)$ of \CRA\ with current and various magnetic fields applied along the $ab$-plane. The resistivity shows a pronounced $\sqrt{T}$-dependence just above \To\ at which $\rho(T)$ deviates from the $\sqrt{T}$ behavior. Data are shifted upwards by 2\,$\mu\Omega\mathrm{cm}$ from the 1\,T curve for better visibility.}
		\label{figS2}
	\end{center}
\end{figure}

\vspace{-0.2cm}
At temperatures just above phases I and II, the resistivity $\rho(T)$ of \CRA\ (with in-plane current) shows a robust $\sqrt{T}$ dependence. This is shown in Fig.~S2. The onset of these phases is then taken as the point at which the temperature dependence deviates from this behavior. For $\mu_0H_{\perp} < 9$\,T, this point corresponds to a kink in $\rho(T)$ whereas for $\mu_0H_{\perp} \geq 9$\,T this point corresponds to the change between the $\sqrt{T}$ and a $T^{2}$ behavior at low temperatures (cf. Fig.~2a of the main article). The process of estimating this point is pictured in Fig.~\ref{figS2}: The square root dependence is fitted (dash lines) and then subtracted from the resistivity. Since the resulting excess resistivity shows a linear temperature dependence below 200\,mK, a linear fit in this range was performed to estimate the transition onset as the zero intersection of the linear fit. The resulting numerical error is then the combination of uncertainty from linear and square root fits. They are plotted as error bars in Fig.~4 of the main article. 
\subsection{AC-susceptibility}
\begin{figure}[ht!]
	\begin{center}
		\includegraphics[width=\columnwidth]{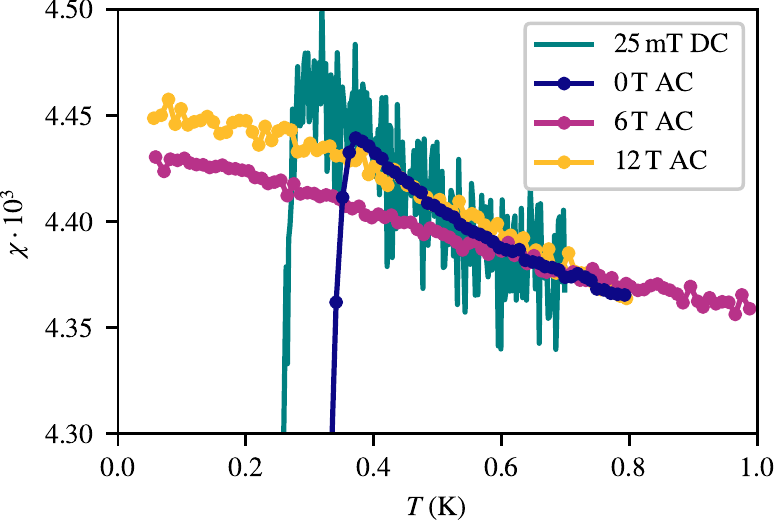}
		\caption{Measured DC- and AC-susceptibility versus temperature at various magnetic fields. For better visibility, 6\,T and 12\,T curves were shifted to match the 0\,T curve at $T$ = 0.8\,K.}
		\label{figS3}
	\end{center}
\end{figure}

\vspace{-0.2cm}
Neither while entering phase I nor phase II there is a substantial signal change in magnetic susceptibility measurements as shown in Fig.~\ref{figS3}. The magnetic susceptibility is given in dimensionless SI units.
\subsection{Pressure dependence of $T_c$ and $T_0$}
\begin{figure}[ht!]
	\begin{center}
		\includegraphics[width=\columnwidth]{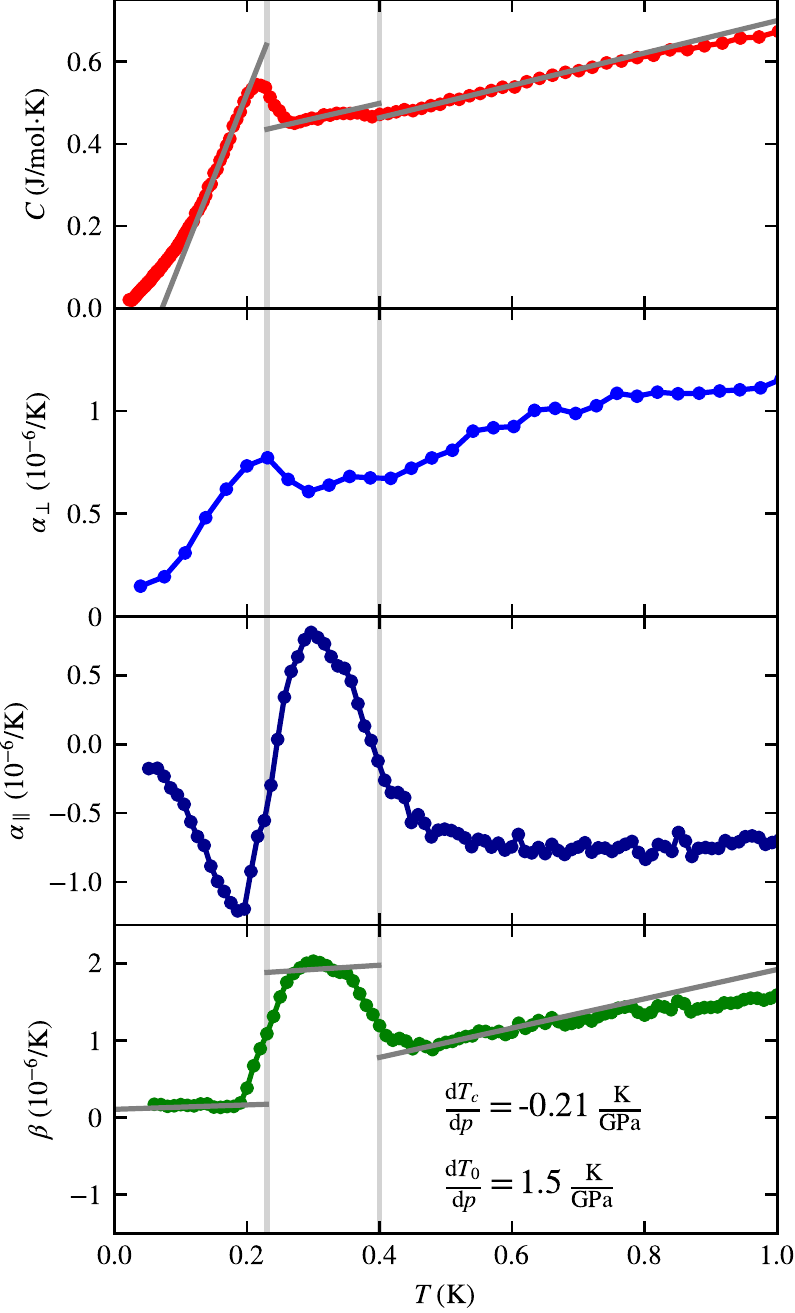}
		\caption{Temperature dependence of specific heat $C$ and thermal expansion coefficient in the plane $\alpha_\perp$, along the c-axis $\alpha_\parallel$ and of the volume $\beta=2\alpha_\perp+\alpha_\parallel$. The gray lines are linear fits to the curves above and below the transitions in order to estimate the jumps in the specific heat and in the thermal expansion at \Tc\ and \To.}
		\label{figS4}
	\end{center}
\end{figure}

\vspace{-0.2cm}
The specific heat and thermal expansion data in Fig.~\ref{figS4} allows to estimate the pressure dependence of $T_c$ and $T_0$ via the Ehrenfest relation for a second order phase transition: $\mathrm{d}T_x/\mathrm{d}p = V_m T_c(\Delta \beta/\Delta C)$. $\Delta C$ and $\Delta\beta$ are the jumps in specific heat and volume thermal expansion at the transition, $V_m$ is the molar volume. The result for both transitions is given in Fig.~\ref{figS4}.
\subsection{Crystalline electric field (CEF) states}
\begin{figure}[ht!]
	\begin{center}
		\includegraphics[width=\columnwidth]{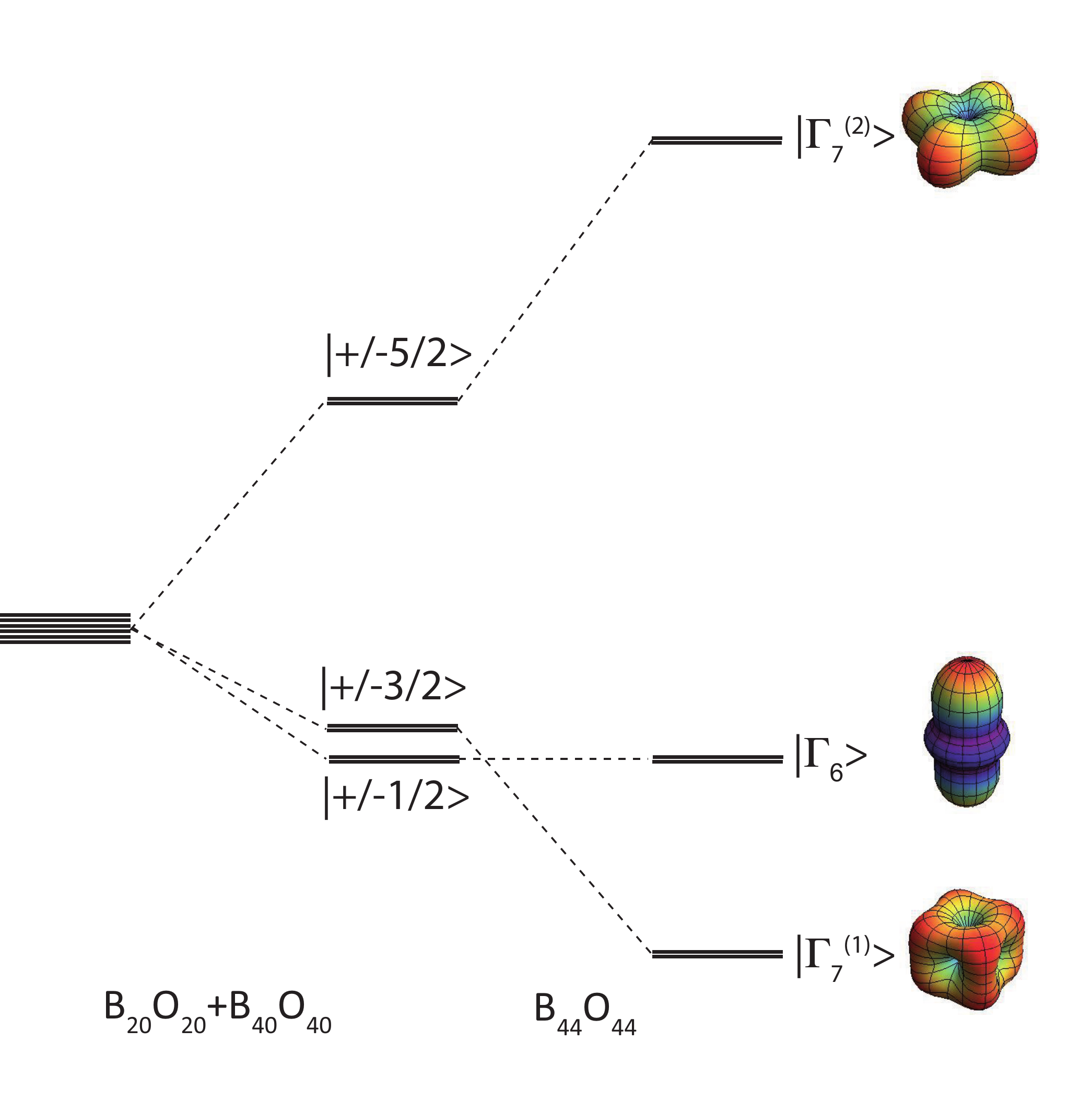}
		\caption{Proposed CEF scheme for \CRA\ which includes how the different contributions $B_{lm}O_{lm}$ of the CEF split the Ce$^{3+}$ $J = 5/2$ multiplet into three Kramers doublets.}
		\label{figS5}
	\end{center}
\end{figure}

\vspace{-0.2cm}
The $J = 5/2$ Hund's rule ground state of Ce$^{3+}$ in \CRA\ is split by the tetragonal crystal field into two $\Gamma_{7}^{(1,2)}$ and one $\Gamma_{6}$ Kramers doublets. For our studies we use the CEF parameters $B_{20}=6.5$\,K, $B_{40}=0.1$\,K and $\left|B_{44}\right|=2.76$\,K which give a decent description of the variation with temperature of the specific heat and the anisotropic magnetic susceptibility. These parameters yield a $\Gamma_{7}$ ground state $\left|\Gamma_{7}^{\left(1\right)},\tau\right\rangle =0.465|\pm5/2\rangle-[\mathrm{sign}B_{44}]0.885|\mp3/2\rangle$, an excited $\left|\Gamma_{6},\tau\right\rangle =|\pm1/2\rangle$ doublet at $\Delta_{1}\simeq30$\,K and an excited $\Gamma_{7}$ state $\left|\Gamma_{7}^{\left(2\right)},\tau\right\rangle =[\mathrm{sign}B_{44}]0.885|\pm5/2\rangle+0.465|\mp3/2\rangle$ at $\Delta_{2}\simeq180$\,K. As the sign of $B_{44}$ is unknown we have considered both cases $B_{44} > 0$ and $B_{44} < 0$. In Fig.~\ref{figS5} we illustrate how the different contributions of the CEF split the $J = 5/2$ state into the 3 doublets. Remarkably, it is the large in-plane contribution $B_{44}O_{44}$ which shifts the $\Gamma_7^{(1)}$ doublet below  the $\Gamma_6$ doublet.
\subsection{Remarks concerning the RBS method}
\vspace{-0.2cm}
The $f$-derived low-energy excitation of a periodic Kondo lattice form coherent quasiparticle. The corresponding Bloch functions are calculated within the Renormalized Band method which combines material-specific ab initio methods and phenomenological considerations in the spirit of the Landau theory of Fermi liquids. For a summary of the method and the underlying ideas, we refer to Refs.~\cite{zwicknagl1992,zwicknagl1993,zwicknagl2011,zwicknagl2016}. Technical details of the calculations will be published in a separate paper. The Renormalized Band method turned out to be a flexible tool with predictive power. An important success was the prediction of the Fermi surface cross sections and anisotropic effective masses of CeRu$_{2}$Si$_{2}$ in 1989~\cite{zwicknagl1990a} that were confirmed experimentally in 1992 by measurements of the dHvA effect \cite{albessard1993,aoki1992,aoki1993}. Further applications were the prediction of the Spin Density Wave (SDW) instability in CeCu$_{2}$Si$_{2}$~\cite{stockert2004} and changes
in the Fermi surface topology (Lifshitz transitions) induced by magnetic fields~\cite{pfau2013,naren2013,pourret2019}. The general shape of the heavy Fermi surface in CeCu$_2$Si$_2$ as predicted in 1993 has been confirmed recently by ARPES experiments~\cite{zwicknagl1993a,wu2021}.

\bibliography{hafner_arXiv_2021}
\bibliographystyle{h-physrev}
\end{document}